\documentclass{ws-ijmpa}

\newcommand\be{\begin{eqnarray}}
\newcommand\ee{\end{eqnarray}}

\begin{document}

\markboth{Kirill Krasnov}
{Gravity as BF theory plus potential}

\catchline{}{}{}{}{}
%
%%%%%%%%%%%%%%%%%%%%%%%%%%%%%%%%%%%%%%%%%%%%%%%%%%%%%%%%%%%%%%%%%%%%

\title{GRAVITY AS BF THEORY PLUS POTENTIAL}

\author{KIRILL KRASNOV}

\address{School of Mathematical Sciences, University of Nottingham, Nottingham, NG7 2RD, UK\\
kirill.krasnov@nottingham.ac.uk}

\maketitle

\begin{history}
\received{Day Month Year}
\revised{Day Month Year}
\end{history}

\begin{abstract}
Spin foam models of quantum gravity are based on Pleba\'nski's formulation of general relativity 
as a constrained BF theory. We give an alternative formulation of gravity
as BF theory plus a certain potential term for the B-field. When the
potential is taken to be infinitely steep one recovers general relativity. For 
a generic potential the theory still describes gravity in that it propagates
just two graviton polarizations. The arising class of theories  
is of the type amenable to spin foam quantization methods, and, we argue,
may allow one to come to terms with renormalization in the spin foam context.

\end{abstract}

\ccode{PACS numbers: 04.60.Pp, 04.50.Kd}

\section{Introduction}	

The starting point of one of the approaches to quantum gravity -- spin foam models -- is 
Pleba\'nski's \cite{Plebanski:1977zz} formulation of general relativity as BF theory supplemented
with the so-called simplicity constraints for the B-field. There are several similar versions of 
this formulation, namely those based on the (complexified) rotation ${\rm SO}(3)$ and Lorentz
${\rm SO}(1,3)$ groups. Currently the most popular is the Lorentz group-based version, see
e.g. \cite{De Pietri:1998mb}, whose advantage is that no additional reality conditions on
the B-field need to be imposed. The new spin foam models of (Riemannian signature) 
quantum gravity, see e.g. \cite{Freidel:2007py}, can be seen to be directly motivated by 
this version of Pleba\'nski theory. 

On the other hand, the original \cite{Plebanski:1977zz} rotation group-based formulation works
with self-dual quantities and therefore, in case one wants to describe metrics of
Lorentzian signature, necessarily involves complex quantities - the price to
pay for its notable algebraic simplicity. However, so far no simple spin foam model 
motivated by the self-dual Pleba\'nski theory was proposed -- even in the
case of Riemannian signature spacetimes where no problems with reality conditions
would exist. See, however, \cite{Reisenberger:1996ib}. This has to do with our
present rather poor understanding of how to translate the self-dual simplicity constraints
into the discrete setting of spin foams. 

The main aim of this contribution is to advertise a certain reformulation of general relativity,
closely related to Pleba\'nski formulation, recently studied in a series of papers by the 
present author. In
this formulation GR is described not as BF theory with constraints, but rather
as BF theory with a potential for the B-field. It is arguably simpler to deal with
potentials than with constraints, both in the continuum or discrete settings.
Thus, the reformulation of GR described in this paper may shed new light on
the problem of quantization of self-dual Pleba\'nski gravity. We also argue
that it may be of help to understand the question of renormalization in the
spin foam context. 

\section{Self-dual Pleba\'nski formulation}

Before we describe a generalization of Pleba\'nski theory that converts it into
a BF theory plus potential let us briefly describe the original, self-dual
formulation \cite{Plebanski:1977zz} of Pleba\'nski with constraints. There are 
some excellent expositions of this formulation, see e.g. \cite{Capovilla:1991qb},
where we send the reader for more details.

The main idea behind Pleba\'nski's formulation of GR is the observation that the
Einstein condition $R_{\mu\nu}\sim g_{\mu\nu}$, where $R_{\mu\nu}$ is the
Ricci tensor of the metric $g_{\mu\nu}$ can be stated as the condition
that the self-dual part of the Riemann curvature tensor $R_{\mu\nu\rho\sigma}$
with respect to first pair of indices is also self-dual with respect to
the second pair. Thus, let us introduce the notion of Hodge operator
as acting on two forms $X_{\mu\nu}$ and sending $X_{\mu\nu}\to X^*_{\mu\nu}=
(1/2)\epsilon_{\mu\nu}^{\quad\rho\sigma} X_{\rho\sigma}$, where $\epsilon_{\mu\nu\rho\sigma}$
is the volume form of the metric $g_{\mu\nu}$, and the indices are raised and
lowered using the metric. Using the Hodge operator one can introduce the projectors
$P^\pm=(1/2)({\rm Id} \pm (1/i) *)$ onto self- and anti-self dual two-forms. 
The Einstein condition can then be checked to be equivalent to, schematically:
\be\label{einstein}
P^+ R P^- =0,
\ee
where $R$ is the Riemann curvature tensor. Let now $A$ be 
the restriction of the Levi-Civita connection for $g_{\mu\nu}$ 
to the bundle of self-dual two-forms. The condition can be stated as
a requirement that the curvature of $A$ is self-dual as a two-form. 
This observation (known in the mathematical literature as the
Atiyah-Hitchin-Singer theorem \cite{Atiyah:1978wi}) is a
way to rephrase Pleba\'nski's formulation \cite{Plebanski:1977zz}.
Interestingly, Pleba\'nski's paper was published one year earlier than \cite{Atiyah:1978wi}.

More concretely, in Pleba\'nski's theory one starts with an ${\rm SO}(3)$
principal bundle $P$ over the spacetime $M$ (complexified ${\rm SO}(3)$ in the case
one wants to describe Lorentzian signature spacetimes). Here $M$ is
a manifold without any additional structure such as a metric on it.
The metric will appear later, only indirectly constructed from other
basic objects of the theory. In physical applications $M$ is non-compact,
and so the bundle $P$ is trivial. Let then $A$ be a connection in the
associated bundle. In addition to the associated bundle, whose fibers are
copies of the Lie algebra ${\mathfrak su}(2)$ of ${\rm SO}(3)$, one can construct
more interesting bundles by tensoring it with the bundles of forms on $M$.
Of particular interest for Pleba\'nski theory is the bundle of Lie-algebra
valued two-forms. Let $B$ be a section of this bundle. This is the
object that will later receive the interpretation of carrying information
about the metric. 

The connection $A$ defines a derivative operator $D_A$ that can be naturally 
extended to act on Lie-algebra-valued forms, in particular on $B$. One
of the basic equations of Pleba\'nski theory then states:
\be\label{pleb-1}
D_A B = 0,
\ee
which can be rephrased by saying that the two-form field $B$ is
``covariantly constant'' with respect to $A$. As it can be shown,
given $B$ this uniquely determines $A$ (provided $B$ satisfies
some non-degeneracy condition).

The second main equation of Pleba\'nski theory states that the curvature
$F=dA+(1/2)A\wedge A$ of $A$, which is a two-form with values
in the Lie algebra of ${\rm SO}(3)$, can be decomposed purely into
the two-forms $B$. Thus, the equation states that there exists
some endomorphism $\Phi$ of the Lie algebra ${\mathfrak su}(2)$
such that:
\be\label{pleb-2}
F = \Phi(B), \qquad \Phi\in {\rm End}_{{\mathfrak su}(2)}.
\ee
To see that this equation is non-vacuous note that a general two-form,
when expanded into a basis of two-forms, has six components. However,
the ${\mathfrak su}(2)$-valued two-form $B$ spans only a three-dimensional
space in the space of two-forms, and so not any two-form is of the form 
(\ref{pleb-2}).

The third equation of Pleba\'nski theory is a condition on the two-form
field $B$, which, in a sense, relates this object to a spacetime metric.
Indeed, being an ${\mathfrak su}(2)$-valued two-form, the field $B$
needs $6\times 3$ numbers to be described. However, a spacetime metric
has only 10 components. At the same time, 3 of the 18 components of
$B$ are ``gauge'' and can be eliminated by the action of ${\rm SO}(3)$.
This means that if $B$ is to describe a spacetime metric it must
satisfy some 5 additional equations. These are as follows. Consider
a four-form $B\wedge_{\otimes} B$. This is a four-form with values in
the symmetric tensor product $({\mathfrak su}(2)\otimes {\mathfrak su}(2))_s$.
However, on $({\mathfrak su}(2)\otimes {\mathfrak su}(2))_s$ we have an
${\rm SO}(3)$-invariant object -- the Killing-Cartan form ${\rm Id}$, or, in other
words, the ${\rm SO}(3)$-invariant metric on ${\mathfrak su}(2)$. A natural
equation on $B$ arises if one requires $B\wedge B$ to be proportional to ${\rm Id}$:
\be\label{pleb-3}
B\wedge_\otimes B \sim {\rm Id},
\ee
where the proportionality coefficient is an arbitrary four-form and can be
determined by taking the trace of this equation. The equation (\ref{pleb-3})
is known as the {\it simplicity} or metricity condition on $B$, and, as
is not hard to see, is actually a set of 5 equations. Thus, it removes 5 out
of the 18 components of $B$, leaving 10 plus the gauge 3, which is the
correct number to describe a spacetime metric. 

The spacetime metric itself arises by observing that a non-degenerate (i.e.
linearly independent) triple of two-forms can be declared to span the
space of {\it self-dual} two-forms, and this defines the notion of
self-duality on two-forms. This is, in turn, equivalent to a conformal
metric, as is shown in e.g. \cite{Samuel}.  What was
shown by Pleba\'nski \cite{Plebanski:1977zz} is that when $B$ satisfies (\ref{pleb-3}) then
the arising metric is unique. Moreover, in this case the connection $A$
satisfying (\ref{pleb-1}) turns out to be just the self-dual part of
the Levi-Civita metric-compatible connection, and then (\ref{pleb-2})
is just the equation (\ref{einstein}) in disguise.

All the above equations can be obtained from a rather simple and natural
action principle. Thus, let us introduce:
\be\label{action}
S[B,A,\Phi]=\int_M {\rm Tr}\left(B\wedge F - \frac{1}{2} B\wedge \Phi(B) \right).
\ee
Here ${\rm Tr}$ is the Killing-Cartan symmetric bilinear pairing on 
${\mathfrak su}(2)$ that above, as an element of 
$({\mathfrak su}^*(2)\otimes {\mathfrak su}^*(2))_s\sim ({\mathfrak su}(2)\otimes {\mathfrak su}(2))_s$
we have denoted by ${\rm Id}$. Varying (\ref{action}) with respect to $A$
one gets (\ref{pleb-1}), varying it with respect to $B$ one gets (\ref{pleb-2}),
and finally varying the action with respect to $\Phi$, which in (\ref{action}) can be seen to
be necessarily symmetric, one obtains (\ref{pleb-3}). Note that to get (\ref{pleb-3}) 
only the traceless part of the matrix $\Phi$ must be varied. The trace part is (proportional to)
the cosmological constant. This finishes our short
description of the Pleba\'nski theory.

In its form (\ref{action}) the theory is that of BF type, i.e. its action
is given by that of $BF$ theory -- the first term in (\ref{action}), plus
an extra non-derivative term for the $B$ field. Such theories are thought
to be susceptible to the spin foam quantization methods. However,
it was so far not possible to understand how the constraints (\ref{pleb-3}) must be
taken care of in the spin foam formalism. For this reason no sufficiently developed
spin foam model motivated by (\ref{action}) exists, see, however, \cite{Reisenberger:1996ib}.

\section{Deformations of general relativity}

As may have become clear from the discussion of the previous section, the
simplicity condition (\ref{pleb-3}) is only used to restrict the possible
two-form fields $B$ considered in the other equations (\ref{pleb-1}), (\ref{pleb-2}).
However, it was also pointed out that an arbitrary two-form field $B$ can be used
to define a conformal metric. Thus, ignoring the issue with the conformal factor for the
moment, one could also consider a gravity theory given by the equations (\ref{pleb-1}), 
(\ref{pleb-2}) without (\ref{pleb-3}) imposed. It turns out that the Bianchi
identities derivable from (\ref{pleb-1}), (\ref{pleb-2}) constrain the system sufficiently
to obtain a closed system of equations, see \cite{Krasnov:2008fm} for a demonstration
of this. Thus, dropping the simplicity constraints (\ref{pleb-3}) one arrives at the
following theory:
\be\label{action-mod}
S[B,A]=\int_M {\rm Tr}(B\wedge F) - \frac{1}{2} V(B\wedge B).
\ee
Here $V(\cdot)$ is an arbitrary gauge-invariant potential function, which,
in order for the action to make sense, is required to be a homogeneous
function of order one in its $({\mathfrak su}(2)\otimes {\mathfrak su}(2))_s$ tensor four-form
valued argument. Indeed, the matrix-valued four-form $B\wedge B$ can always be
written as some four-form times a matrix, and then the four-form can be
pulled out of the potential function using its homogeneity. The result
can then be integrated over $M$ to produce a scalar. The homogeneity
of $V$ then guarantees that this does not depend on which four-form is
used. 

The field equations following from (\ref{action-mod}) are essentially
unchanged from (\ref{pleb-1}), (\ref{pleb-2}). Thus, when varying the
action with respect to $A$ one gets (\ref{pleb-1}). When varying
the action with respect to the two-form field one gets (\ref{pleb-2}),
with the only novelty being that the endomorphism $\Phi$ that in Pleba\'nski theory
was left unspecified, is now determined by the two-form field itself, via
the potential $V$. Thus, we get:
\be
F = \frac{1}{2} \frac{\partial V}{\partial B},
\ee
where this equation makes sense in view of the homogeneity of the potential.

The action (\ref{action-mod}) defines a much larger class of theories
than (\ref{action}), for the potential $V(\cdot)$ may be quite
arbitrary. In particular, the new class of theories is parametrized
by an infinite number of the ``coupling constants'' -- e.g. the coefficients
of the decomposition of $V(\cdot)$ into a power series. It is not hard to 
show, see e.g. \cite{Krasnov:2008fm} for details, that when the
potential $V(\cdot)$ becomes infinitely steep (in an appropriate sense)
then one effectively imposes (\ref{pleb-3}) and recovers GR. For
a generic potential the theory (\ref{action-mod}) is, however, 
distinct from general relativity. In spite of this, it can be shown
quite easily, see e.g. \cite{Krasnov:2007cq}, that the theory in
question still propagates just two graviton polarizations, 
exactly as in GR. Thus, the class of theories (\ref{action-mod})
provides an infinite parameter family of {\it deformations} of GR,
with the Einstein's theory easily recoverable from (\ref{action-mod})
via a simple limit. The question of how to recover a preferred, physical metric from
the conformal class of metrics naturally described by (\ref{action-mod}) has also
been recently resolved in \cite{Krasnov:2008ui}.

The infinite-parameter class of theories (\ref{action-mod}) was first written down
in \cite{Bengtsson:1990qg}, generalizing an earlier work by Capovilla  (later published as 
\cite{Capovilla:1992ep}) that studied a one-parameter family of deformations dubbed 
"neighbors of GR" by the author. Both \cite{Bengtsson:1990qg} and \cite{Capovilla:1992ep}
describe deformations using the so-called pure connection formulation of 
GR \cite{Capovilla:1991kx}. Thus, it would be rather hard to recognize (\ref{action-mod}) 
in the results of \cite{Bengtsson:1990qg} and \cite{Capovilla:1992ep}. The same theory was 
later rediscovered in \cite{Krasnov:2006du} starting directly from the Pleba\'nski formulation. 
The equivalence between the two descriptions was demonstrated in \cite{Bengtsson:2007zzd}. 
The formulation described here (with a potential for the two-form field) 
is spelled out in more details in e.g. \cite{Krasnov:2008fm}.

\section{Quantum Gravity}

We would like to argue that it is the class (\ref{action-mod}) -- not general
relativity (\ref{action}) -- that should be used as a natural starting point for the spin
foam (and possibly perturbative) quantization of gravity. 

First, the theory in its version (\ref{action-mod}) does not contain
any additional ``Lagrange multiplier'' fields such as $\Phi$ in (\ref{action}).
For this reason it is likely to be much easier to deal with in the discrete
spin foam setting, where we already have a good deal of experience with
working with the discrete versions of the $B$ and connection fields. 
Thus, it seems likely that the class of theories (\ref{action-mod})
is amenable to the spin foam quantization techniques rather directly.
It just remains to be determined how to translate the property of
the potential being a homogeneous function of degree one into the
discrete setting.

The second point we would like to make may initially sound a bit unconventional 
from the spin foam perspective. The point is that Einstein's theory, given in Pleba\'nski formulation
by (\ref{action}), is non-renormalizable. Thus, one does not expect this theory
to play any direct role in the Planckian regime that we would like our
quantum gravity to describe. If anything, one should expect that the relevant
Planck scale theory should be a modified one, taking into account the
quantum corrections that become essential at high energies. Such a much
larger class of theories that takes into account at least some (if not
all) the possible corrections is given by (\ref{action-mod}). Indeed, its
original motivation in \cite{Krasnov:2006du} was precisely to come to
terms with the renormalization in quantum gravity.

Translated into the language of spin foams this renormalization motivation
may be formulated as follows. In spin foam approach to quantum gravity one
obtains an amplitude for a manifold by ``gluing'' together amplitudes
for the individual spacetime simplices, see e.g. \cite{Freidel:2007py}
and references therein for more details. Let us consider the ``renormalization''
in the context of spin foams, i.e. analyze what happens when one computes the
simplex $\sigma$ amplitude as the result of integration over the labels of the
``smaller" simplices that are glued together to make $\sigma$ 
(in a technical jargon this corresponds to an e.g. $5\to 1$ move).
When the elementary simplex amplitudes are built as dictated by the Pleba\'nski action (\ref{action})
(or its ${\rm SO}(4)$ version), the new simplex amplitude -- the result of the spin foam 
``renormalization group flow'' -- is of a different type, not anymore
describable as coming from the original Pleba\'nski action. This is,
we believe, how the non-renormalizability of GR manifests itself
in the spin foam context. Thus, the spin foam renormalization group
flow does not preserve the classical action (\ref{action}) one starts from.
As we have already said, we find this entirely natural, and having to
do with the non-renormalizability of the underlying theory.

It is however possible (but quite non-trivial to show) that some larger
class of theories may be closed under such a renormalization group flow. In
the discrete setting of spin foams this would manifest itself in the
simplex amplitude given by the result of the $5\to 1$ move being of
the same type as one started from, but with all the coupling constants
-- parameters of the theory -- being changed in some subtle way. 
Should one find the class of theories with such a property, one can
then see whether its UV completion exists by determining whether
there is some non-trivial UV fixed point of the flow. This fixed point, if 
exists, would then provide the sought UV theory. It is then clear that the first
step in the direction of this program is to enlarge the class
of gravity theories that is being considered. We would like to
propose the class (\ref{action-mod}) as a viable and natural arena
for these ideas in the spin foam context.

\section*{Acknowledgments}

The author was supported by an EPSRC Advanced Fellowship.

\end{document}